\documentclass[a4paper]{jpconf}

\usepackage{graphicx}

\begin{document}

\title{Recent results of high-energy spin phenomena of gluons and sea-quarks in polarized proton-proton collisions at RHIC at BNL}

\author{Bernd Surrow}

\address{Temple University \\ 
Department of Physics \\
1900 N. 13th Street \\ 
Philadelphia, PA 19122 \\
USA}
\ead{surrow@temple.edu}

\begin{abstract}

The STAR experiment at the Relativistic Heavy-Ion Collider at Brookhaven National
Laboratory is carrying out a spin physics program in high-energy polarized proton 
collisions at $\sqrt{s}=200\,$GeV and $\sqrt{s}=500\,$GeV to gain a deeper insight 
into the spin structure and dynamics of the proton. 

One of the main objectives of the spin physics program at RHIC is the precise 
determination of the polarized gluon distribution function. The STAR detector is 
well suited for the reconstruction of various final states involving jets, 
$\pi^{0}$, $\pi^{\pm}$, e$^{\pm}$ and $\gamma$, which allows to measure several 
different processes. Recent results suggest a gluon spin contribution to the proton 
spin at the same level as the quark spin contribution itself.

The production of $W$ bosons in polarized p+p collisions at $\sqrt{s}=500\,$GeV
opens a new era in the study of the spin-flavor structure of the proton. $W^{-(+)}$ 
bosons are produced in $\bar{u}+d\;(\bar{d}+u)$ collisions and can be detected 
through their leptonic decays, $e^{-}+\bar{\nu}_{e}\;(e^{+}+\nu_{e})$, where only 
the respective charged lepton is measured. Results of $W^{-(+)}$ production suggest 
a large asymmetry between the polarization of anti-$u$ and anti-$d$ quarks.

\end{abstract}

\section{Introduction}	

High energy polarized $p+p$ collisions at a center-of-mass energy of 
$\sqrt{s}=200\,$GeV and at $\sqrt{s}=500\,$GeV at RHIC at BNL provide a unique way to probe the proton spin 
structure using very well established processes in high-energy
physics, both experimentally and theoretically.
The polarized gluon and quark /
antiquark measurements using jet and W boson production at RHIC have been 
formulated as key milestones of the
US Nuclear Physics program. A recent global analysis provides
for the first time evidence
of a non-zero value of the gluon polarization 
$\int_{\tiny 0.05}^{\tiny 0.2}\Delta g \, dx \,(Q^{2}=10\,{\rm GeV}^{2}) = 0.1^{+0.06}_{-0.07}$,
i.e. of similar magnitude as
the total quark spin contribution to the proton spin \cite{Aschenauer:2013woa}. 
The analysis of data
in 2009 using for the
first time collisions of polarized protons at $\sqrt{s}=500\,$GeV
established a novel scheme to probe the quark flavor structure
using $W$ boson production. Data taken in 2012 suggest a large
asymmetry of the spin contribution of anti-u quarks and anti-d quarks
similar to the large and well-known difference between the momentum
distributions of anti-u quarks and anti-d quarks. Highlights of these results 
will be presented in the following two sections.

\section{High-energy spin phenomena of gluons} 

\begin{figure}[t]
\centerline{\includegraphics[width=140mm]{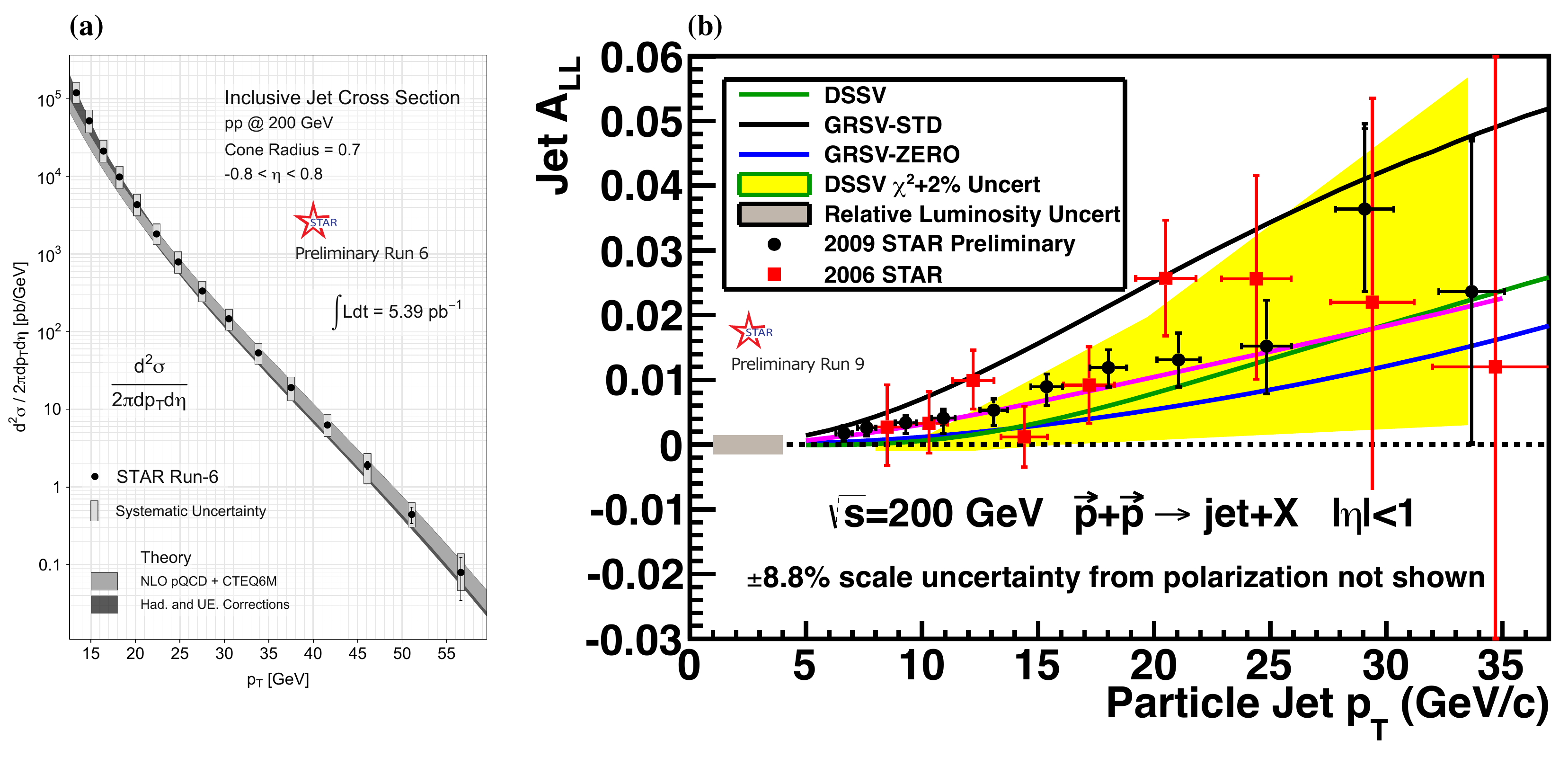}}
\caption{\label{Fig-Inclusive-Jets}{{\it Cross-section based on STAR Run 6 data (a) and 
longitudinal double-spin asymmetry based on STAR Run 9 data (b) for inclusive jet production 
as a function of $p_{T}$ in comparison 
to NLO calculations.}}}
\end{figure}

The gluon polarization program is based on the measurement of various probes, such
as inclusive jet production and di-jet production in polarized proton-proton
collisions at RHIC using the STAR experiment. The full
exploitation of the STAR Barrel and Endcap Electromagnetic
Calorimeter (BEMC / EEMC) is crucial for
this analysis effort.  
Both the first measurement at STAR of the
longitudinal double-spin asymmetry for inclusive jet production
\cite{Abelev:2006uq} and updates
with higher precision \cite{Abelev:2007vt,Adamczyk:2012qj} along 
with the first neutral pion measurement \cite{Abelev:2009pb} 
have been published. Precision
measurements of inclusive
jet production \cite{collaboration:2011fga} along with the first significant 
di-jet measurements \cite{Walker:2011vs}
have been presented suggesting a statistically non-zero gluon spin
contribution to the proton spin at the level of $\int_{\tiny 0.05}^{\tiny 0.2}\Delta g \, dx \,(Q^{2}=10\,{\rm GeV}^{2}) = 0.1^{+0.06}_{-0.07}$,
i.e. of similar magnitude as the total quark spin contribution to the
proton spin.

\begin{figure}[t]
\centerline{\includegraphics[width=140mm]{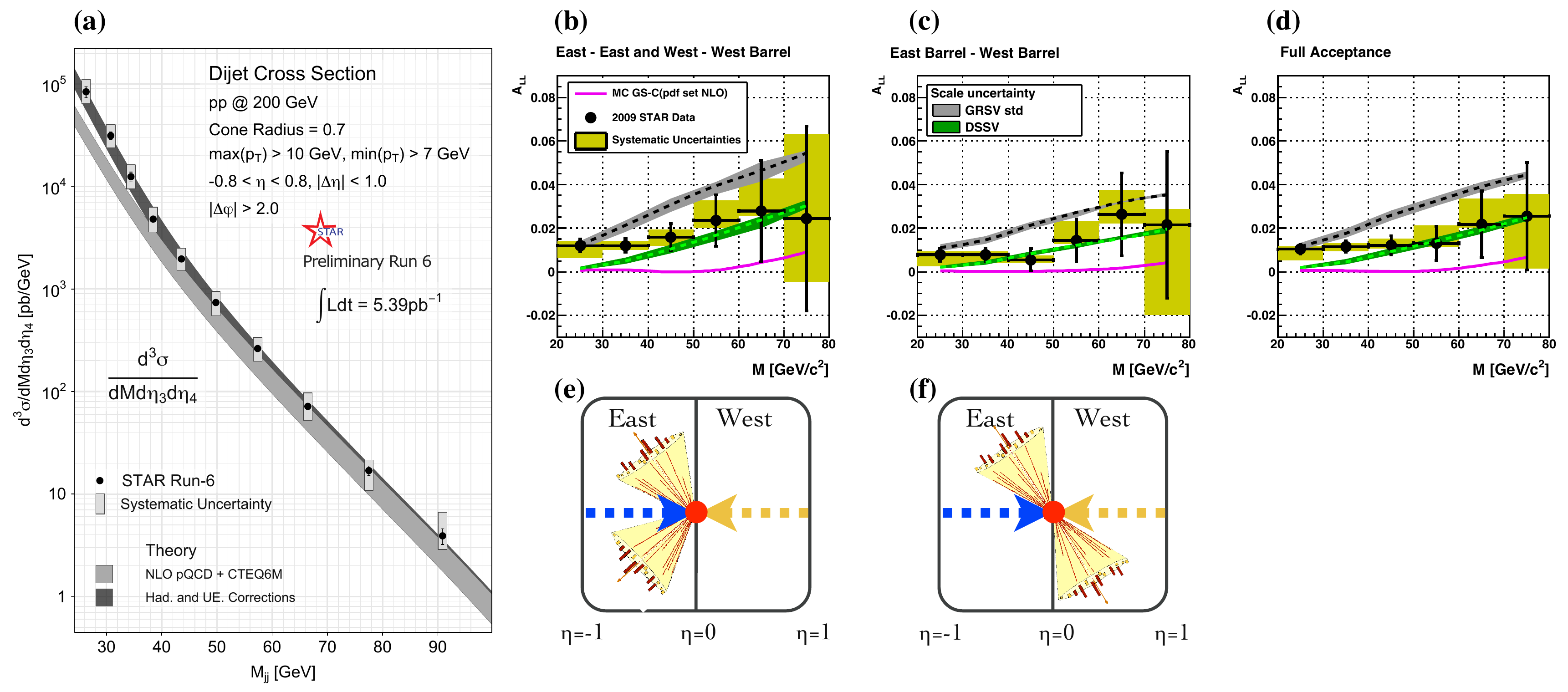}}
\caption{\label{Fig-Di-Jets}{{\it Cross-section (a) and longitudinal double-spin asymmetry for 
three different $\eta$ regions (b), (c) and (d) for di-jet production as a function of $M$ 
based on Run 9 STAR data in comparison to NLO calculations. Illustration of asymmetric (e) and 
symmetric (f) collisions.}}}
\end{figure}

\noindent The measurement of the gluon polarization through inclusive measurements such as
jet production and $\pi^{0}$ and $\pi^{\pm}$ production was the
prime focus of the physics analysis program of the Run 3/4 and Run 5 data
samples as well as the Run 6 and Run 9 data samples. 
The sensitivity of those
inclusive measurements to the underlying gluon polarization in high-energy
polarized proton-proton collisions has been discussed in detail in
\cite{Bunce:2000uv,Jager:2004jh,Surrow:2003aj}.

\noindent The 2006 inclusive jet cross-section measurement is shown in
Figure~\ref{Fig-Inclusive-Jets} (a) \cite{Surrow:2010zz}.
Good agreement is found between data and full NLO calculations taking into account corrections for
underlying event and hadronization effects.  
Figure \ref{Fig-Inclusive-Jets} (b) shows the most recent STAR preliminary result of $A_{LL}$ for inclusive jet production as a function
of $p_{T}$ based on the 2006 and 2009 data sample. The 2006 analysis has
been published \cite{Adamczyk:2012qj}. The 2009 data fall in-between the
GRSV-STD ($\Delta G=\int_{\tiny 0}^{\tiny 1 }\Delta g \, dx
\,(Q^{2}=10\,{\rm GeV}^{2}) = 0.4$) \cite{Gluck:2000dy} and 
DSSV ($\Delta G = \int_{\tiny 0}^{\tiny 1 }\Delta g \, dx \, (Q^{2}=10\,{\rm GeV}^{2})=0.0$) 
\cite{deFlorian:2008mr} NLO calculations for $A_{LL}$. The DSSV fit
result was constrained by the 2005 \cite{Abelev:2007vt} and 2006 \cite{Adamczyk:2012qj}
data. The 2009 result suggests a larger gluon polarization than the one
suggested by DSSV, driven by the low-$p_{T}$ region. Prior to a full
global analysis incorporating the 2009 result, a first step to quantify the impact of the 2009
result has been made by W. Vogelsang et al. \cite{deFlorian:2011ia}. This study suggests a
non-zero gluon polarization at the level of 
$\int_{\tiny 0.05}^{\tiny 0.2}\Delta g \, dx \, (Q^{2}=10\,{\rm GeV}^{2})= 0.13$
and thus of similar magnitude as the quark spin contribution
to the proton spin. 

\begin{figure}[t]
\centerline{\includegraphics[width=160mm]{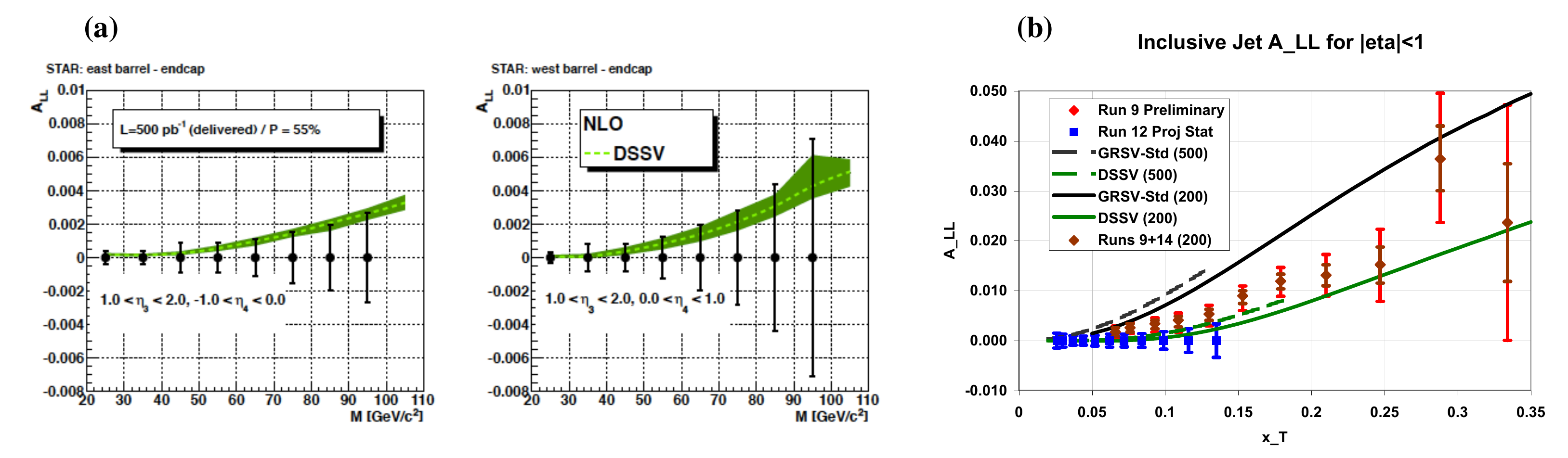}}
\caption{\label{Fig-Projections-ALL}{{\it Projection of STAR longitudinal double-spin asymmetry measurements for di-jet (a) and inclusive jet (b) production.}}}
\end{figure}

\noindent Di-jet production at STAR will allow a
better constraint of the underlying event kinematics to extract
the shape of the gluon polarization in a global analysis
\cite{Stratmann:2006ix}.  The invariant di-jet mass is proportional to the sum
of the partonic Bjorken-x  values, ${\rm x_{1}+x_{2}}$, whereas the pseudo-rapidity sum
${\rm \eta_{3}+\eta_{4}}$ is proportional to the logarithmic ratio
of the partonic Bjorken-x values, ${\rm \log(x_{1}/x_{2})}$. 
Measurements at both ${\rm
\sqrt{s}=200\,}$GeV and ${\rm \sqrt{s}=500\,}$GeV are preferred to maximize
the kinematic region in ${\rm x}$.
The wide acceptance of the STAR experiment permits reconstruction of
di-jet events with different topological configurations,
i.e. different $\eta_{3}$/$\eta_{4}$ combinations, ranging from
symmetric ($x_{1}=x_{2}$) partonic collisions to asymmetric
($x_{1}<x_{2}$ or $x_{1}>x_{2}$) partonic 
collisions (Figure~\ref{Fig-Di-Jets} (e) and (f)). This, together
with the variation of the center-of-mass energy, is expected to constrain $\Delta g$
over a wide range in $x$ of approximately $\simeq 3\cdot 10^{-3} < x <
0.3$.  The NLO framework for correlation measurements exists and
therefore those measurements can be used in a global
analysis~\cite{deFlorian:1998qp}. The needed extension from the first
di-jet measurement in 2009 at mid-rapidity to the forward rapidity
region requires an extension of jet reconstruction into the STAR EEMC
acceptance region ($1<|\eta|<2$). The utilization of the STAR Forward
GEM Tracker to provide tracking information in addition to the
calorimetric information provided by the STAR EEMC will be
important. 
Figure \ref{Fig-Di-Jets} (a) shows the first di-jet
cross-section measurement \cite{Sakuma:2010ms}. Good agreement
is found between data and full NLO calculations taking into account corrections for
underlying event and hadronization effects. 
Figures~\ref{Fig-Di-Jets} (b) - (d) show the first significant measurement of the 
longitudinal double-spin asymmetry, $A_{LL}$, as a function of the
di-jet invariant mass, $M$, for different pseudo-rapidity regions (\ref{Fig-Di-Jets} (e) and (f)) 
emphasizing different Bjorken-$x$ regions. The 2009 di-jet $A_{LL}$ measurement also suggests a
larger gluon polarization scenario than those suggested by 
DSSV \cite{deFlorian:2008mr}. The
data fall in-between the GRSV-STD \cite{Gluck:2000dy} and 
DSSV \cite{deFlorian:2008mr} polarized PDFs 
used for a full NLO 
calculation similar to the inclusive-jet result shown in Figure
\ref{Fig-Inclusive-Jets}. 

\begin{figure}[t]
\centerline{\includegraphics[width=140mm]{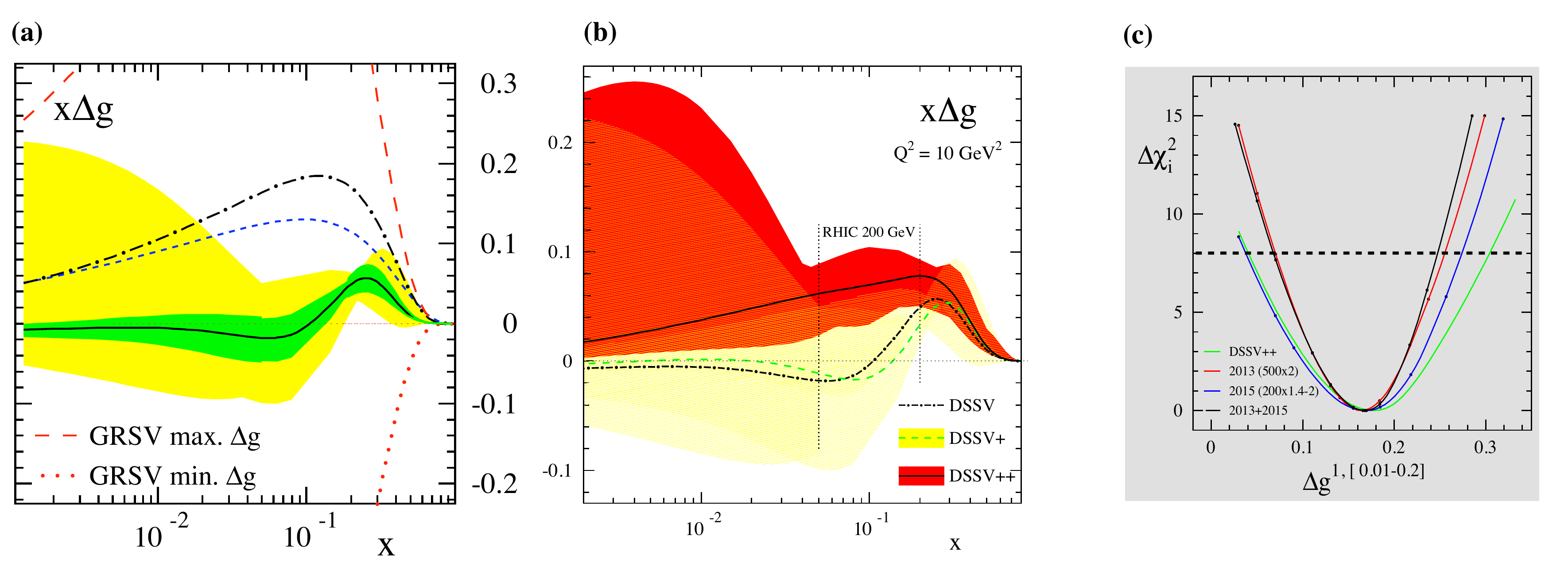}}
\caption{\label{Fig-Projections-deltaG}{{\it Impact of the Run 9 
STAR inclusive jet and PHENIX neutral pion results on the polarized gluon 
distribution function (a) and (b) together with the $\Delta \chi^{2}$ 
distribution as a function (c) of the truncated first moment of $\Delta g$ 
restricted to $0.05<x<0.2$. The achieved precision is improved by
approximately a factor of 2 compared to the original DSSV uncertainty 
band \cite{deFlorian:2008mr}.}}}
\end{figure}

\noindent All current PHENIX and STAR ${\rm A_{LL}}$ measurements 
taken together, in comparison to different NLO perturbative QCD predictions
for ${\rm A_{LL}}$, provide a consistent picture which rules out a large gluon
polarization scenario. At the same time, data taken in Run 9 provide
the first evidence for a non-zero gluon spin contribution to the
proton spin.
A critical aspect to extract the
gluon polarization of the proton is to perform a global analysis of several
${\rm A_{LL}}$ measurements obtained from the PHENIX and STAR
collaborations, taking into account a constraint of polarized PDFs 
at high Bjorken-x values by several polarized
fixed-target DIS experiments. 

Figure~\ref{Fig-Projections-deltaG} shows the impact of
the Run 9 and future data sets on the polarized gluon distribution
function \cite{Aschenauer:2013woa}. The Run 9 result suggests a statistically non-zero
value of the gluon polarization at the level of $\int_{\tiny
  0.05}^{\tiny 0.2}\Delta g \, dx \,(Q^{2}=10\,{\rm GeV}^{2}) = 0.1^{+0.06}_{-0.07}$ which is of
similar magnitude as the total quark spin contribution itself. Future measurements will focus on extending the range
towards smaller values in Bjorken-$x$ and providing higher precision
measurements by approximately a factor of 2 compared to the original DSSV
fit uncertainties \cite{deFlorian:2008mr}.

\section{High-energy spin phenomena of sea-quarks}

\begin{figure}[t]
\centerline{\includegraphics[width=130mm]{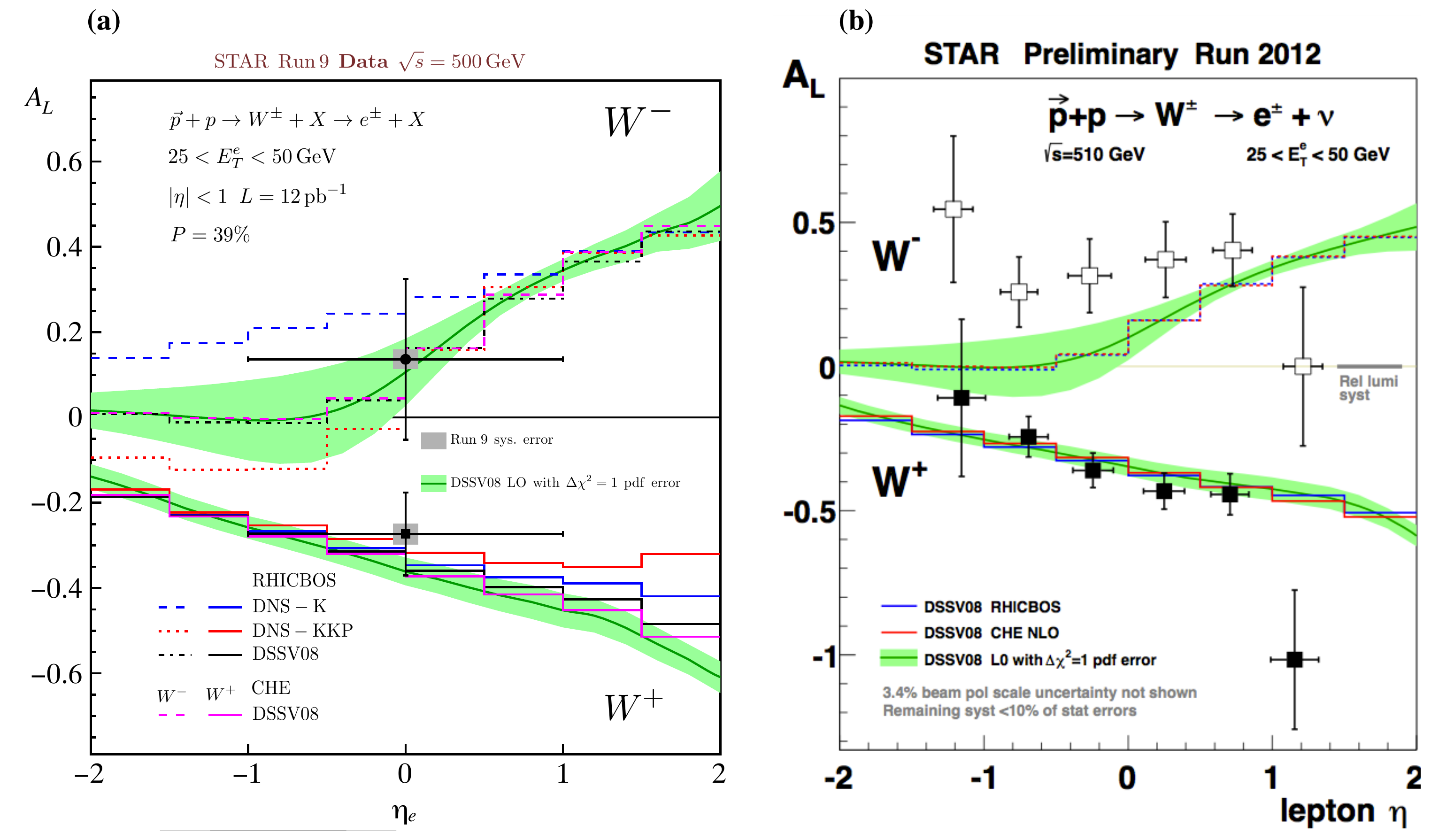}}
\caption{\label{Fig-W}{{\it First STAR longitudinal single-spin asymmetry for $W^{\pm}$ production based on Run 9 data (a) and preliminary results from Run 12 (b).}}}
\end{figure}

\begin{figure}[t]
\centerline{\includegraphics[width=140mm]{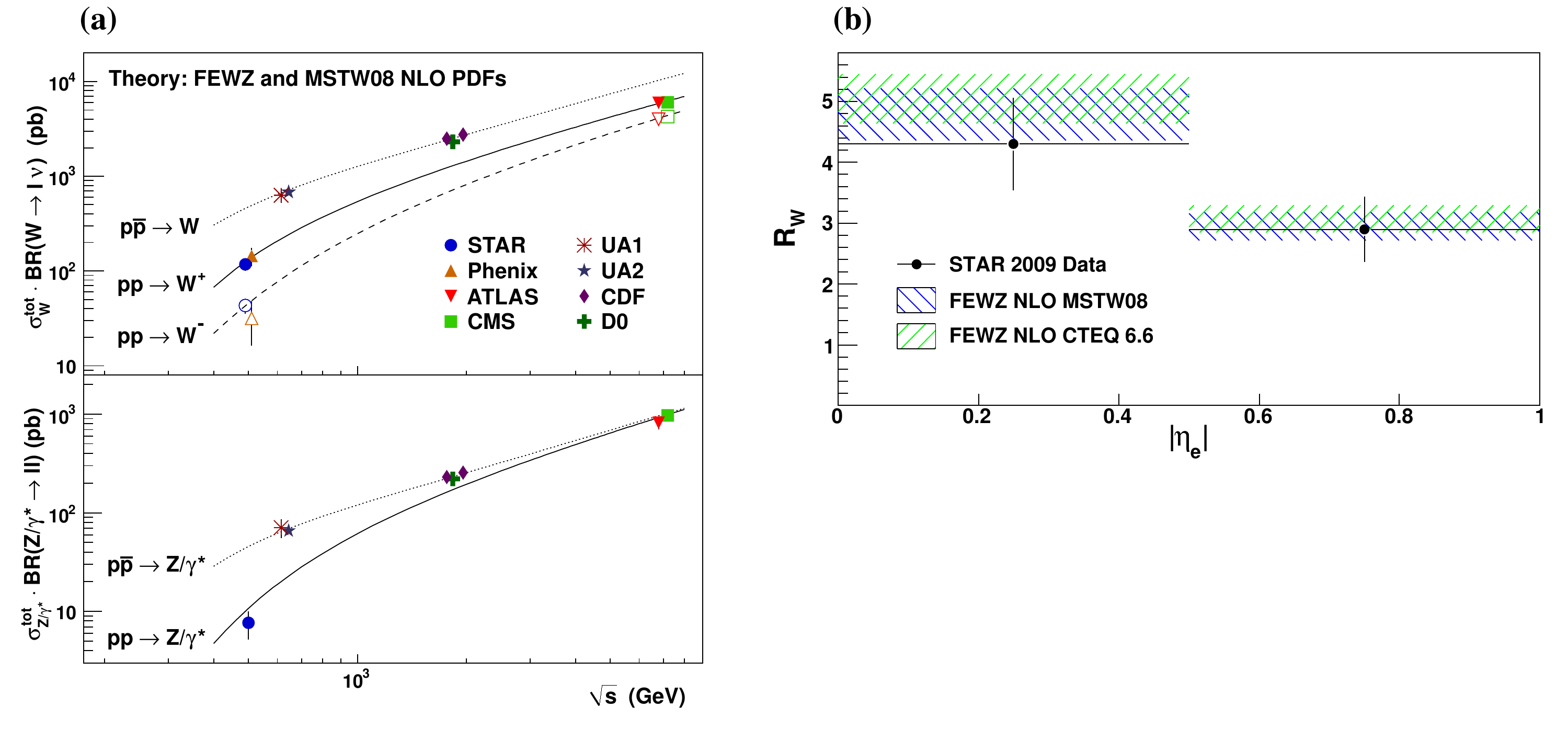}}
\caption{\label{Fig-W-Cross}{{\it $W/Z$ cross-section measurements as a function of $\sqrt{s}$ (a) and the $W^{+}/W^{-}$ cross-section ratio measurement  as a function of $|\eta|$ (b).}}}
\end{figure}

The data taking period in 2009 of polarized $p+p$ collisions at $\sqrt{s}=500\,$GeV opened a new era in the 
study of the spin-flavor structure of the proton based on the
production of $W^{-(+)}$ bosons. 
$W^{-(+)}$ bosons are
produced predominantly through $\bar{u}+d$ $(u+\bar{d})$ collisions and can be detected through
their leptonic decay \cite{Bourrely:1993dd}. 
Quark and antiquark polarized parton-distribution functions (PDFs) are probed in calculable leptonic $W$ decays at large
scales set by the mass of the $W$ boson.
The production of $W$ bosons in polarized proton collisions allows for the observation of purely 
weak interactions, giving rise to large, parity-violating, longitudinal single-spin asymmetries.
A theoretical framework has been developed to describe inclusive lepton ($l$) 
production, $p+p \rightarrow W^{\pm}+X \rightarrow l^{\pm}+X$, that can be directly compared with experimental
measurements using constraints on the transverse energy and pseudorapidity of the 
final-state leptons \cite{Nadolsky:2003ga,deFlorian:2010aa}.
This development profits from a rich history of hadroproduction of weak bosons 
at the CERN SPS and the FNAL Tevatron and provides a firm basis to use
$W$ production as a new high-energy probe in polarized $p+p$ collisions \cite{Kotwal:2008zz}.
The subsequent analyses of both the W boson
asymmetry and cross-section measurements have been published \cite{Aggarwal:2010vc,STAR:2011aa}.
The first cross-section measurement is shown in Figure \ref{Fig-W-Cross}. 
Data taken in 2012 suggest a large
asymmetry of the spin contribution of anti-u quarks and anti-d quarks
similar to the large and well-known asymmetry between the momentum
distributions of anti-u quarks and anti-d quarks \cite{Stevens:2013raa}.

\begin{figure}[t]
\centerline{\includegraphics[width=80mm]{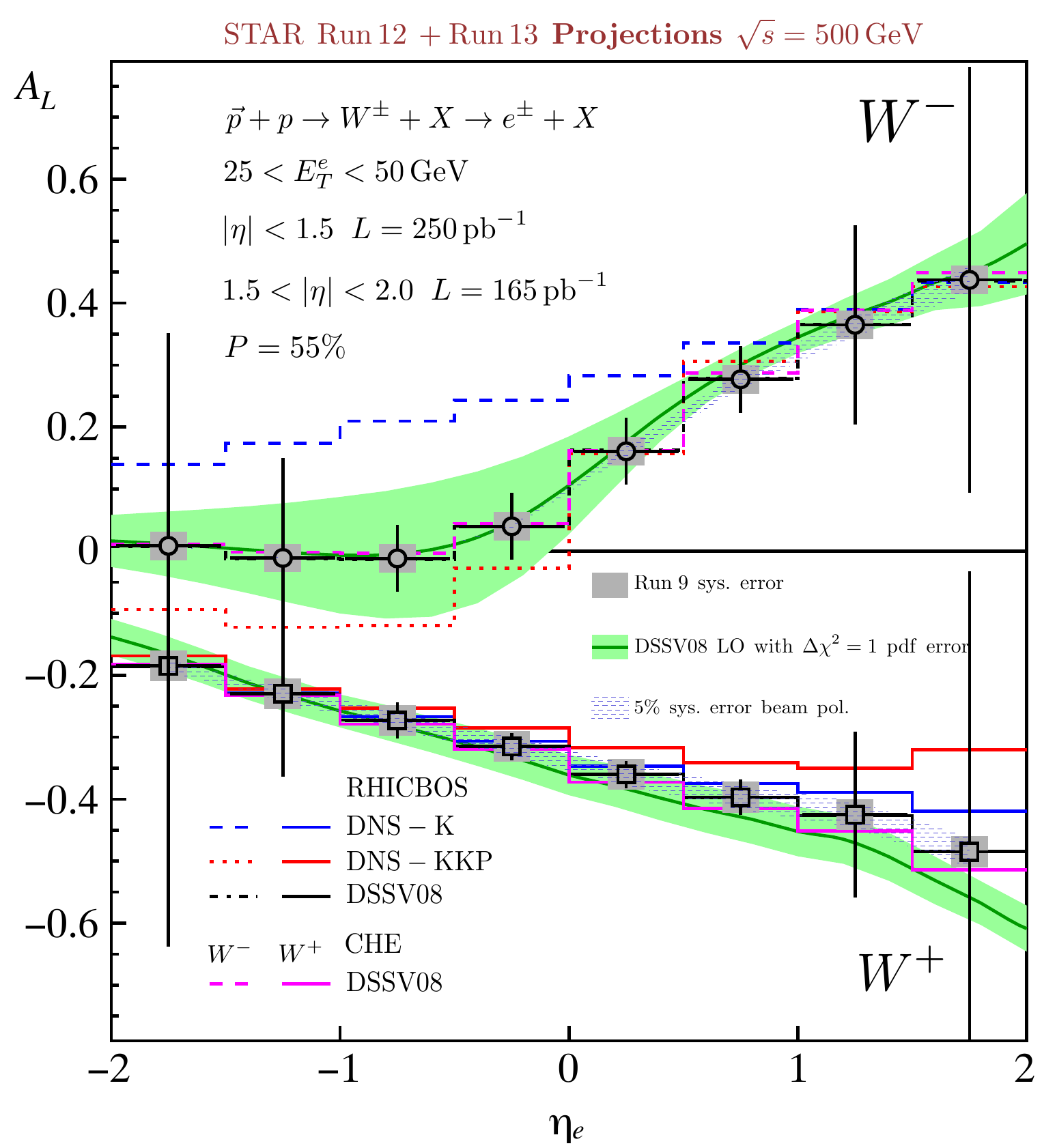}}
\caption{\label{Fig-Projections-AL}{{\it Projected longitudinal single-spin asymmetry for $W$ production as a function of $\eta$.}}}
\end{figure}

\noindent An upgrade of the STAR forward tracking system was carried out
with the installation of the
Forward GEM Tracker (FGT) in 2012 prior to the 2013 data taking period
\cite{Surrow:2010zza}. 
This upgrade will improve the ability of the STAR tracking system to 
track high-energetic
electrons (positrons) from $W^{-(+)}$-boson decays at forward
acceptance enhancing the sensitivity to quark / antiquark polarizations.

The first published $W$ measurement at mid-rapidity
using the Run 9 data sample is based on $L=12\,$pb$^{-1}$ \cite{Aggarwal:2010vc}.
This achievements marks a major milestone of the RHIC Spin program, in
particular the ability of the STAR detector system to establish the
reconstruction of $W$ bosons, both for a cross-section and asymmetry measurement.
The STAR detector systems used in this analysis are the STAR Time Projection Chamber (TPC), the STAR Barrel Electromagnetic Calorimeter (BEMC) 
and the STAR Electromagnetic Endcap Calorimeter (EEMC). Only their tower response has been taken into
account. 
The BEMC was used to measure the transverse energy, $E_{T}$, of $e^\pm$. 
The suppression of QCD background by several orders of magnitude was based on the TPC, BEMC and EEMC.
The STAR TPC plays a critical role 
to provide an efficient way for high-$p_{T}$ charge-sign
discrimination of electron / positron track candidates. 
The reconstruction of $W$ boson events is well
established including the treatment of background. This will be the
basis for future measurements, in particular the large Run 13 data
sample of $L\simeq 300\,$pb$^{-1}$.  The critical new step is the development of
tracking software concerning the charge-sign separation of
high-$p_{T}$ electrons (positrons) from $W^{-(+)}$-decays 
for $1<|\eta|<2$ using the STAR FGT.

\noindent Figure~\ref{Fig-W} (a) shows the measured leptonic asymmetries $A^{W^+}_{L}$
and $A^{W^-}_{L}$ for $|\eta_{e}|<1$ and $25<E^{e}_{T}<50\,$GeV from Run 9. The vertical black error bars include only the statistical uncertainties. The systematic
uncertainties are indicated as grey bands. The statistical uncertainties dominate over the systematic uncertainties.
The measured asymmetries are compared to predictions based on full resummed (\textsc{rhicbos}) 
\cite{Nadolsky:2003ga} and NLO (\textsc{che}) \cite{deFlorian:2010aa} calculations. The \textsc{che} calculations use the DSSV08 
polarized PDFs \cite{deFlorian:2008mr}, whereas the \textsc{rhicbos} 
calculations are shown in addition for the older DNS-K and DNS-KKP \cite{deFlorian:2005mw} PDFs.  The \textsc{che} 
and \textsc{rhicbos} 
results are in 
good agreement. The range spanned by the DNS-K and DNS-KKP distributions for $\Delta \bar{d}$ and $\Delta \bar{u}$ coincides, 
approximately, with the corresponding DSSV08 uncertainty estimates \cite{deFlorian:2008mr,deFlorian:2009vb}.
The spread of predictions for $A^{W^{+(-)}}_{L}$ is largest at forward (backward) $\eta_{e}$ and is strongly correlated to the 
one found for the $\bar{d}$ ($\bar{u}$) polarized PDFs in the RHIC kinematic region in contrast to the backward (forward)
$\eta_{e}$ region dominated by the behavior of the well-known valence $u$ ($d$) polarized PDFs \cite{deFlorian:2010aa}.
At midrapidity, $W^{+(-)}$ production probes a combination of the polarization of the $u$ and $\bar{d}$ ($d$ and $\bar{u}$) quarks, 
and $A^{W^{+(-)}}_{L}$ is expected to be negative (positive)
\cite{deFlorian:2008mr,deFlorian:2009vb}.  Any measurements beyond
this first result will focus on making $\eta_{e}$ dependent
asymmetry measurements at midrapidity together with measurements at forward and 
backward pseudorapidities constraining the polarization of $\bar{d}$ and $\bar{u}$  quarks.

\noindent A major step forward has been achieved with the release of preliminary
results based on the Run 12 data sample shown in Figure~\ref{Fig-W}
(b) \cite{Stevens:2013raa}. While the $W^{+}$ asymmetry result is consistent with the DSSV 
prediction \cite{deFlorian:2008mr}, the
$W^{-}$ asymmetry result suggests a larger $\bar{u}$  quark
polarization. This is indeed the finding from a full global analysis
incorporating the Run 12 asymmetry results. 

\begin{figure}[t]
\centerline{\includegraphics[width=150mm]{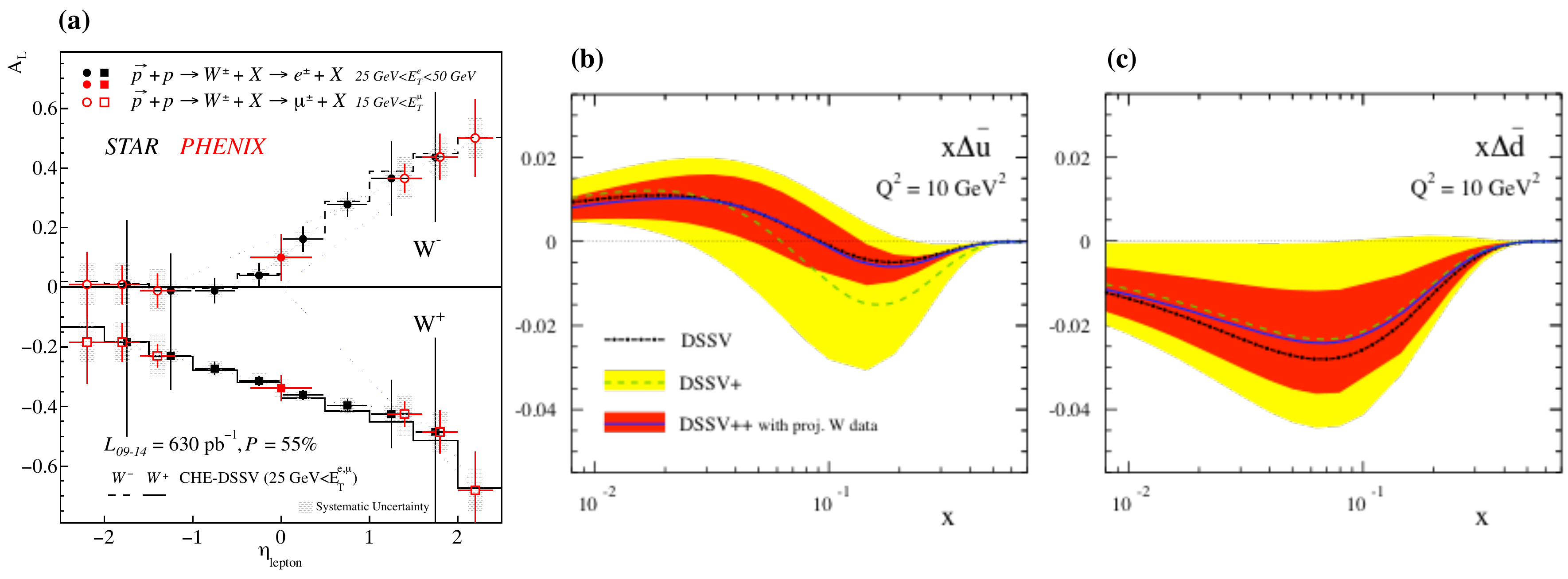}}
\caption{\label{Fig-Projections-deltaq}{{\it Impact of RHIC $W$ measurements 
of $A_{L}$ (a) on antiquark polarizations, $\Delta \bar{u}$ (b) and $\Delta \bar{d}$ 
(c). The red shaded band is approximately a factor of 2 smaller compared to 
the original DSSV uncertainty band 
\cite{deFlorian:2008mr, deFlorian:2009vb}.}}}
\end{figure}

Figure~\ref{Fig-Projections-deltaq} shows the impact on the polarized 
$u$ and $d$ antiquark distribution functions \cite{Aschenauer:2013woa}. Future RHIC
$W$ asymmetry measurements shown in
Figure~\ref{Fig-Projections-deltaq} (a) will have a substantial impact
in comparison to a global fit result which does not include any
$W$ asymmetry results from RHIC by approximately a factor of 2 compared to the
original DSSV fit uncertainties \cite{deFlorian:2008mr,deFlorian:2009vb} 
(Figure ~\ref{Fig-Projections-deltaq} (b), (c)). The impact of the Run 12 preliminary
results suggest a large difference of $x(\Delta \bar{u} - \Delta \bar{d})$ of about 0.07. This result is intriguing pointing to a
broken QCD sea, similar to the unpolarized case.  
Extending these measurements to backward and forward pseudo-rapidity
regions has been a long-standing goal of the RHIC Spin program and is now even
more important with the recent finding in Run 12.

\section{Summary}

In summary, new results of jet production at RHIC in polaried proton-proton 
collision suggest a gluon spin contribution of similar magnitude as the quark 
spin contribution itself. It is the hope that these findings can be confronted 
with Lattice QCD calculations. The ground work has been laid by X. Ji et al. in
a recent paper \cite{Ji:2013fga,Ji:2013dva,X-Ji-QCD}. Results of $W$ boson production suggest a large asymmetry 
in the  QCD sea, i.e. the difference of the polarization of anti-u and anti-d 
quarks. Such a scenario has been discussed in the past in the context of a valence 
scenario by the GRSV group \cite{Gluck:2000dy} and in 
particular by J. Soffer \cite{Bourrely:2001du,Bourrely:2013qfa}. Recent lattice QCD calculations presented 
by H. Lin in \cite{Huey-Wen-Lin} point to the same finding. The experimental goal of the RHIC Spin 
program is an extension of the kinematic region at low-$x$ along with an 
increase of precision for both di-jet production and $W$ boson production in 
particular in the backward / forward pseudo-rapidity region

\section*{References}

% \bibliographystyle{h-physrev3}
% \bibliography{surrow}

\begin{thebibliography}{0}    %for 1 digit

\bibitem{Aschenauer:2013woa}
E.~Aschenauer {\em et~al.},
(2013), arXiv/1304.0079.
%%CITATION = ARXIV:1304.0079;%%

\bibitem{Abelev:2006uq}
STAR Collaboration, B.~Abelev {\em et~al.},
{\it Phys. Rev. Lett.} {\bf 97}, 252001 (2006).

\bibitem{Abelev:2007vt}
STAR Collaboration, B.~Abelev {\em et~al.},
{\it Phys. Rev. Lett.} {\bf 100}, 232003 (2008).

\bibitem{Adamczyk:2012qj}
STAR Collaboration, L.~Adamczyk {\em et~al.},
{\it Phys. Rev.} {\bf D86}, 032006 (2012).

\bibitem{Abelev:2009pb}
STAR Collaboration, B.~Abelev {\em et~al.},
{\it Phys. Rev.} {\bf D80}, 111108 (2009).

\bibitem{collaboration:2011fga}
STAR Collaboration, P.~Djawotho,
(2011), arXiv/1106.5769.
%%CITATION = ARXIV:1106.5769;%%

\bibitem{Walker:2011vs}
STAR Collaboration, M.~Walker,
(2011), arXiv/1107.0917.
%%CITATION = ARXIV:1107.0917;%%

\bibitem{Bunce:2000uv}
G.~Bunce, N.~Saito, J.~Soffer, and W.~Vogelsang,
{\it Ann. Rev. Nucl. Part. Sci.} {\bf 50}, 525 (2000).

\bibitem{Jager:2004jh}
B.~Jager, M.~Stratmann, and W.~Vogelsang,
{\it Phys. Rev.} {\bf D70}, 034010 (2004).

\bibitem{Surrow:2003aj}
STAR Collaboration, B.~Surrow,
{\it AIP Conf. Proc.} {\bf 675}, 318 (2003).
%%CITATION = APCPC,675,318;%%

\bibitem{Surrow:2010zz}
STAR Collaboration, B.~Surrow,
{\it PoS - DIS 2010 Conference (Florence, Italy)} {\bf 1}, 249 (2010).
%%CITATION = POSCI,DIS2010,249;%%

\bibitem{Gluck:2000dy}
M.~Gluck, E.~Reya, M.~Stratmann, and W.~Vogelsang,
{\it Phys. Rev.} {\bf D63}, 094005 (2001).

\bibitem{deFlorian:2008mr}
D.~de~Florian, R.~Sassot, M.~Stratmann, and W.~Vogelsang,
{\it Phys. Rev. Lett.} {\bf 101}, 072001 (2008).

\bibitem{deFlorian:2011ia}
D.~de~Florian, R.~Sassot, M.~Stratmann, and W.~Vogelsang,
{\it Prog. Part. Nucl. Phys.} {\bf 67}, 251 (2012).

\bibitem{Stratmann:2006ix}
M.~Stratmann,
{\it Word Scientific - DIS 2006 Conference (Tsukuba, Japan)} {\bf 1}, 715
  (2006).
%%CITATION = INSPIRE-738385;%%

\bibitem{deFlorian:1998qp}
D.~de~Florian, S.~Frixione, A.~Signer, and W.~Vogelsang,
{\it Nucl. Phys.} {\bf B539}, 455 (1999).

\bibitem{Sakuma:2010ms}
STAR collaboration, T.~Sakuma and M.~Walker,
{\it J. Phys. Conf. Ser.} {\bf 295}, 012068 (2011).

\bibitem{Bourrely:1993dd}
C.~Bourrely and J.~Soffer,
{\it Phys. Lett.} {\bf B314}, 132 (1993).
%%CITATION = PHLTA,B314,132;%%

\bibitem{Nadolsky:2003ga}
P.~M. Nadolsky and C.~Yuan,
{\it Nucl. Phys.} {\bf B666}, 31 (2003).

\bibitem{deFlorian:2010aa}
D.~de~Florian and W.~Vogelsang,
{\it Phys. Rev.} {\bf D81}, 094020 (2010).

\bibitem{Kotwal:2008zz}
A.~V. Kotwal and J.~Stark,
{\it Ann. Rev. Nucl. Part. Sci.} {\bf 58}, 147 (2008).
%%CITATION = ARNUA,58,147;%%

\bibitem{Aggarwal:2010vc}
STAR Collaboration, M.~Aggarwal {\em et~al.},
{\it Phys. Rev. Lett.} {\bf 106}, 062002 (2011).

\bibitem{STAR:2011aa}
STAR Collaboration, L.~Adamczyk {\em et~al.},
{\it Phys. Rev.} {\bf D85}, 092010 (2012).

\bibitem{Stevens:2013raa}
STAR Collaboration, J.~R. Stevens,
(2013), arXiv/1302.6639.
%%CITATION = ARXIV:1302.6639;%%

\bibitem{Surrow:2010zza}
B.~Surrow,
{\it Nucl. Instrum. Meth.} {\bf A617}, 196 (2010).
%%CITATION = NUIMA,A617,196;%%

\bibitem{deFlorian:2005mw}
D.~de~Florian, G.~Navarro, and R.~Sassot,
{\it Phys. Rev.} {\bf D71}, 094018 (2005).

\bibitem{deFlorian:2009vb}
D.~de~Florian, R.~Sassot, M.~Stratmann, and W.~Vogelsang,
{\it Phys. Rev.} {\bf D80}, 034030 (2009).

\bibitem{Ji:2013fga}
X.~Ji, J.-H. Zhang, and Y.~Zhao,
{\it Phys. Rev. Lett.} {\bf 111}, 112002 (2013).
%%CITATION = ARXIV:1304.6708;%%

\bibitem{Ji:2013dva}
X.~Ji, (2013), arXiv/1305.1539.
%%CITATION = ARXIV:1305.1539;%%

\bibitem{X-Ji-QCD}
X.~Ji,
Proceedings of the QCD Evolution Workshop (QCD2013), Newport News,
  VA.

\bibitem{Bourrely:2001du}
C.~Bourrely, J.~Soffer, and F.~Buccella,
{\it Eur.Phys.J.} {\bf C23}, 487 (2002).
%%CITATION = HEP-PH/0109160;%%

\bibitem{Bourrely:2013qfa}
C.~Bourrely, F.~Buccella, and J.~Soffer,
(2013), arXiv/1308.3567.
%%CITATION = ARXIV:1308.3567;%%

\bibitem{Huey-Wen-Lin}
H.-W. Lin,
Proceedings of the QCD Evolution Workshop (QCD2013), Newport News,
  VA.

\end{thebibliography}

\end{document}